\documentclass[aps]{revtex4}
\usepackage{amsfonts}
\usepackage{amsmath}
\usepackage{amssymb}
\usepackage{graphicx}
\usepackage{bbold}

\usepackage[colorlinks, citecolor=blue,linkcolor=blue]{hyperref}

\begin{document}
 \title{Large Distance Continuous Variable Communication with Concatenated Swaps}
\author{Muhammad Asjad, Stefano Zippilli, Paolo Tombesi, and David Vitali}
 \affiliation{School of Science and Technology, Physics Division, University of Camerino, Camerino (MC), Italy}
\begin{abstract}
The radiation-pressure interaction between electromagnetic fields and mechanical resonators can be used to efficiently entangle two light fields coupled to the same mechanical mode. We analyze the performance of this process under realistic conditions, and we determine the effectiveness of the resulting entanglement
as a resource for quantum teleportation of continuous-variable light signals over large distances, mediated by concatenated swap operations. We study the sensitiveness of the protocol to the quality factor of the mechanical systems, and its performance in non-ideal situations in which losses and reduced detection efficiencies are taken into account.
\end{abstract}
\maketitle


\section{Introduction}

Entanglement is the characteristic trait of quantum mechanics and is a fundamental resource for many quantum technology applications. In particular, quantum  information and communication exploit entanglement to enhance the efficiency and the security for the manipulation of information.
Quantum key distribution~\cite{Eckert,Yamamoto}, quantum teleportation ~\cite{Bennett,Braunstein&Kimble} and dense coding~\cite{Bennett2,Braunstein&Kimble2} are few exemplary protocols that take advantage of quantum mechanics to achieve optimal performances.
In this context the idea of a quantum internet has been recently envisaged~\cite{kimble}, where quantum
connection between distant nodes is established in order to perform quantum information protocols between distant sites~\cite{Aspel,zeilinger,cinesi}, which could also involve Earth-satellites communications~\cite{space}. For this purpose it seems necessary to use quantum repeaters or at least faithful quantum relays~\cite{zoller,gisin}, that are devices able to redistribute entangled resources created locally or between close locations, to larger distances.

Most efficient strategies to transfer information between distant locations involve light. In the continuous variable domain~\cite{Eisert2003,Braunstein2005,Paris2005,Adesso2007,Weedbrook20012} two-mode squeezed light is the prominent resource of many quantum communication schemes. Recently cavity optomechanics has emerged as a versatile platform for the manipulation and control of electromagnetic fields both at optical and microwave frequencies~\cite{AMO,rmp}. Ground breaking experiments have achieved, for example, cooling to the quantum ground state of a nanomechanical resonator~\cite{oconnell,teufel,chan}, generation of squeezed light at the cavity output~\cite{brooks,painter,regal}, coherent conversion between optical and mechanical degrees of freedom~\cite{verhagen,palomaki}, and entanglement between optical and mechanical degrees of freedom~\cite{Lenhert}. Even though it has not been experimentally achieved yet, cavity optomechanical setups have been also proposed as a viable alternative to second-order nonlinear optical parametric amplifiers for the generation of electromagnetic two-mode squeezed fields~\cite{giovannetti,claudiu&mari}. The effective parametric interaction mediated by the optomechanical coupling with a nanomechanical resonator has the drawback of being very sensitive to the thermal noise affecting the mechanical degree of freedom, but has the unique feature of allowing the generation of two-mode squeezing between fields at \emph{completely different} frequencies such as microwave and optical ones~\cite{shabir1,lin,clerk,Barzanjeh}. First experiments in this direction have already demonstrated the transfer and readout of signals at completely different wavelengths~\cite{hill,bochmann,bagci,lehnert14}.

In this article we propose a scheme for achieving long-distance communication with continuous variables~\cite{Grosshans&Grangier} specifically exploiting entanglement between radiation at different wavelengths generated with cavity optomechanics. The entangled beams can be combined via an entanglement swapping protocol~\cite{zukowski,pan,xia,takei,lloyd,mehdi1,mehdi2,mehdi3} realizing a quantum relay; by suitably concatenating a number of such optomechanical devices one can create a long-distance continuous variable entangled link which could be used, for example, for teleporting with good fidelity an unknown coherent state at extremely large distances, even in the presence of losses and non-unit detection efficiency. In particular, the optomechanical device at a node could generate entanglement between a near-infrared field matching a specific transparency window of the atmosphere (say 810 nm) and a field at telecom wavelengths (say 1550 nm). By performing the homodyne Bell measurements on the near-infrared beams propagating in free space up to distant satellite stations, entanglement could be swapped to the telecom wavelength fields at distances much longer than those allowed by standard telecom fiber loss. By replicating such scheme one could extend such continuous variable entangled link even further. The present scheme still exponentially decreases with distance due to the effect of loss and to the lack of quantum memories; nonetheless, it could represent a starting point for a quantum repeater scheme because the mechanical mode at each optomechanical node could be potentially used as a good quantum memory due to its very long damping time. This latter fact represents a further motivation for exploiting cavity optomechanics rather than standard parametric amplifiers for entangling the radiation modes.

The paper is organized as follows. In Section II we shortly introduce the model and review some known results. In Section III, we consider two optomechanical entanglement sources and study the possibility to extend the reach of the generated entanglement using an entanglement swapping protocol. Section IV is devoted to the analysis of concatenated swaps operations over a chain of optomechanical entanglement sources. The effects of photon losses and finite detection efficiency are considered in Section V. Finally, in Section VI we draw our conclusions and present further considerations.

\section{The Model}\label{model}

We consider a Fabry-Perot cavity with an oscillating end mirror~\cite{AMO}.
Two modes of the cavity, at frequencies $\omega_a$ (mode A) and $\omega_b$ (mode B) are driven, respectively, by two optical fields of frequencies $\omega_{La}/2\pi$,
and $\omega_{Lb}/2\pi$ with powers $P_a$ and $P_b$.
The perfectly reflecting end mirror of the optical cavity oscillates at
frequency $\omega_m/2\pi$ along the cavity axis,
and couples to the electromagnetic fields with strengths $g_j= \sqrt{\hbar /m\omega_m}(\omega_j/L)\,$, where the index $j=a,b$ distinguishes the two light modes, $L$ is the equilibrium cavity length, and $m$ is the effective mass of the end mirror. Moreover, the cavity modes can lose photons at rates $\kappa_j$ and the mirror dissipates mechanical energy at rate $\gamma$ in a thermal bath a temperature $T$, and with $n_{th}$ average thermal phonons.
The corresponding Hamiltonian is given by \cite{AMO,giovannetti},
\begin{eqnarray}
\hat{H}&=&\hbar\omega_a \hat{a}^\dagger \hat{a}+\hbar\omega_b \hat{b}^\dagger \hat{b}+\dfrac{\hbar\omega_m}{2}(p^2 + q^2) - \left(g_a \hat{a}^\dagger \hat{a}+g_b  \hat{b}^\dagger \hat{b} \right)\hat{q}\nonumber\\
&&+i\hbar\left(E_{a}\hat{a}^\dagger e^{-i\omega_{La} t} + E_{b}\hat{b}^\dagger e^{-i\omega_{Lb} t}-H.c.\right),\label{hml}
\end{eqnarray}
where, $\hat{a} (\hat{b})$ and $\hat{a}^\dagger (\hat{b}^\dagger)$ are the annihilation and creation operators for the optical cavity modes fulfilling the bosonic commutation relations
 $[\hat{a},\hat{a}^\dagger]=[\hat{b},\hat{b}^\dagger]=1$,
while $\hat{q}$ and $\hat{p}$ denote the dimensionless position and momentum operators of the mechanical mode satisfying $[\hat{q},\hat{p}]=i$.
Finally, the last line in Eq.~(\ref{hml}) accounts for the external driving of the cavity modes where the strength $E_a$ ($E_b$) is related to the input laser power $P_a$ ($P_b$) by the relation
$|E_a|=\sqrt{2P_a\kappa_a/\hbar\omega_{La}}$ ($|E_b|=\sqrt{2P_b\kappa_b/\hbar\omega_{Lb}}$).
The optomechanical dynamics is in general non-linear, but it can be linearized around the mean-field stationary solution when the driving power is sufficiently large. Then, the steady state of the resulting Gaussian linear dynamics can be easily solved in the Fourier domain.

\paragraph*{Entanglement between output optical modes.}

We are interested in the entanglement between two optical modes at the output of the opto-mechanical cavity. According to the input-output theory~\cite{Collett&Gardiner}, the operators for the output fields are related to the cavity and to the input noise operators by the relation
\begin{equation}
\hat a_{out}=\sqrt{2\kappa}\hat a -\hat a_{in}, \ \ \ \ \ \ \ \hat b_{out}=\sqrt{2\kappa}\hat b -\hat b_{in},
\end{equation}
where the only non-zero correlation functions of the input operators are $\langle\hat a_{in}(t) \hat a^\dagger_{in}(t')\rangle=\langle\hat b_{in}(t)\hat b^\dagger_{in}(t')\rangle=\delta(t-t')$~\cite{Gardiner&Zoller}.
The entanglement of the output modes can be analyzed in terms of temporal modes, appropriately filtered from the total field with specific filter functions~\cite{AMO,Zippilli14}. A simple and straightforward example is the one-pole filter which is expressed
in time and Fourier domains as
\begin{equation}\label{filter}
f_j(t)=\sqrt{\dfrac{2}{\tau_j}} e^{-(\dfrac{1}{\tau_j}+i\Omega_j)t} \theta(t),\quad \,f_j(\omega)=\dfrac{\sqrt{\tau_j/\pi}}{1+i\tau_j(\Omega_j-\omega)},\quad j=a,b,
\end{equation}
where $\theta(t)$ is the Heaviside step function, and $\Omega_j$  and $1/\tau_j$ are, respectively, the central frequency and the bandwidth of the $j$-th filter.
The corresponding filtered fields are described by the causal bosonic annihilation operators
\begin{equation}
\hat a_{out,\Omega_a}(t)=\int^{t}_{-\infty} f_a(t-t')\hat a_{out}(t') dt',\quad \hat b_{out,\Omega_b}(t)=\int^{t}_{-\infty} f_b(t-t')\hat b_{out}(t') dt'. \label{out}
\end{equation}
Since the system is Gaussian, all the information about the output light is contained in the first and second order moments of the field operators.
In particular, it is convenient to introduce the quadratures $\hat X_{out},\hat Y_{out}$ of the filtered output modes, that are defined by the relation $\hat\alpha_{out,\Omega_{\alpha}}=(\hat X_{out,{\alpha}}+i \hat Y_{out,{\alpha}})/\sqrt{2}$ with ${\hat \alpha}=\hat a,\hat b$, and fulfill the commutation relations $[\hat X_{out,\alpha} , \hat Y_{out,\alpha}]=i$. Hence, the
  entanglement properties of the output modes can be extracted by the analysis of the corresponding
covariance matrix, whose elements are defined as
\begin{equation}
({\bf V}_{out} (t))_{i,j}=\dfrac{\langle\hat R_{out,i}\hat R_{out,j}+\hat R_{out,j}\hat R_{out,i} \rangle }{2} -\langle\hat R_{out,i}\rangle \langle\hat R_{out,j}\rangle,
\end{equation}
where ${\bf\hat  R}_{out}(t)=[\hat X_{out,a},\hat Y_{out,a},\hat X_{out,b},\hat Y_{out,b}]^{T}$ is the column vector of the filtered output mode quadratures, with
the superscript $T$ indicating transposition.

The stationary solution for the covariance matrix ${\bf V}_{out}$ can be readily obtained by employing the strategy discussed in Ref.~\cite{AMO}. Specifically it can be expressed
as
\begin{equation}
{\bf V}_{out}=\int^{\infty}_{-\infty} {\cal T}(\omega)\left[(i\omega+{\cal A})^{-1}+{\cal P}\right]{\cal D}\left[(-i\omega+{\cal A})^{-1}+{\cal P}\right]^{T} {\cal T}(-\omega)^{T}\ d\omega, \label{vout}
\end{equation}
where ${\cal P}$ is the diagonal matrix ${\cal P}=Diag[0,0,1/2\kappa_a,1/2\kappa_a,1/2\kappa_b,1/2\kappa_b]$, ${\cal A}$ is the matrix of coefficients of the system of quantum Langevin equations for the field quadratures and the position and momentum of the mechanical oscillator and it is defined as
\begin{equation}
{\cal A}=\left( \begin{array}{cccccc}
0        &-\omega_m    &0          &0          &0            &0          \\
-\omega_m &-\gamma_m    &g_a        &0          &g_b  &0        \\
0        &0            &-\kappa_a  &\Delta_a   &0            &0         \\
g_a      &0            &-\Delta_a  &-\kappa_a  &0            &0         \\
0        &0            &0          &0          &-\kappa_b    &\Delta_b      \\
g_b      &0            &0          &0          &-\Delta_b    &-\kappa_b
\end{array}\right), \label{Afb}
\end{equation}
${\cal D}$ is the corresponding diffusion matrix given by
${\cal D}=Diag[0,\gamma_m (2n_{th}+1),\kappa_a, \kappa_a, \kappa_b, \kappa_b]$, and finally
${\cal T}(\omega)$ is a $4\times 6$ matrix which contains the filter functions and is defined as
\begin{equation}
{\cal T}(\omega)=\frac{1}{\sqrt{2}}\left( \begin{array}{cccccccc}
0   &0   &\sqrt{\kappa_a} \tilde h_a^+(\omega)&-\sqrt{\kappa_a} \tilde h_a^+(\omega)&0     \\
0   &0        &\sqrt{\kappa_a} \tilde h_a^-(\omega)&\sqrt{\kappa_a} \tilde h_a^-(\omega) &0 \\
0    &0   &0   &0       &\sqrt{\kappa_b} \tilde h_b^+(\omega)&-\sqrt{\kappa_b} \tilde h_b^-(\omega)\\
0     &0        &0   &0       &\sqrt{\kappa_b} \tilde h_b^-(\omega)&\sqrt{\kappa_b}\tilde h_b^+(\omega)\end{array}
\right),
\end{equation}
where $\tilde h_j^\pm(\omega)=\tilde f_j(\omega)\pm\tilde f_j(-\omega)^*$ with $\tilde f_j(\omega)$ defined in Eq.~(\ref{filter}) \cite{AMO,Barzanjeh}. We note that the matrix ${\cal T}(\omega)$ is constructed as a non-square matrix in order to select only the output field operators, so that ${\bf V}_{out}$ is, indeed, a $4\times 4$ matrix for the output field correlations.
We remark that the system is stable and can reach a steady state only if all the eigenvalues of the  matrix ${\cal A}$ have negative real part. In general, explicit stability conditions can be obtained by applying the Routh-Hurwitz criterion \cite{routh}.
In the following we describe the corresponding entanglement that is generated in the steady state of realistic optomechanical devices (hence, corresponding to the regime of optomechanical stability), and we discuss its dependence on the quality factor  $Q_m=\omega_m/\gamma$ of the mechanical resonator.

\paragraph*{Stationary output entanglement between mode A and mode B:}

It is convenient to express
the $4\times 4$ covariance matrix  of the reduced Gaussian state of the two filtered output fields ${\bf V}_{out}$, in terms of 2x2 matrices (whose specific form can be evaluated following the procedure discussed in Ref.~\cite{AMO}) as
\begin{equation}
{\bf V}_{out}=\left( \begin{array}{cc}
{\bf A}    &{\bf D}      \\
{\bf D}^T    &{\bf B}
\end{array}\right), \label{cov}
\end{equation}
with $A$ and $B$ symmetric.
Here we measure the corresponding entanglement in terms of the logarithmic negativity~$E_{N}$~\cite{Vidal&Werner,eisert,plenio}, which, for Gaussian states is  defined as~\cite{illuminati}
\begin{equation}
E_{N}=\max(0,-\ln\,2\eta_-), \label{EN}
\end{equation}
where $\eta_-$ is the smallest symplectic eigenvalue of the partially transposed covariance matrix, given by
\begin{equation}\label{eta}
\eta_-=2^{-\frac{1}{2}}\left [ \Sigma({\bf V}_{out})-\sqrt{\Sigma({\bf V}_{out})^2-4Det({\bf V}_{out})} \right]^{\frac{1}{2}},
\end{equation}
with
\begin{equation}
\Sigma({\bf V}_{out})=Det({\bf A})+Det({\bf B})-2Det({\bf D}).
\end{equation}
\begin{figure}
\includegraphics[scale=1]{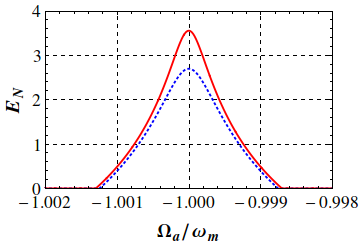}
\caption{(Color online) Logarithmic negativity $E_N$ plotted as a function of the central frequency $\Omega_a$ of the filter of mode A, for the following values of the Q-factor, $Q_m=10^5$ (blue dotted curve) and $Q_m=10^7$ (red curve). The other system parameters are $\omega_m=2\pi\times 10 $ MHz, $L=0.001 $ m, $\kappa_a=\kappa_b=0.2 \omega_m$, $T=4.2 $ K, $\omega_m \tau_b=\omega_m \tau_a=300$, $\lambda_a=1550$ nm, $\lambda_b=810$ nm, $P_a=17$ mW, $P_b=6$ mW, and $\Omega_{b}=\omega_m$. Moreover, as a necessary condition for entanglement, we choose opposite detunings $\Delta_a=\omega_m$ and $\Delta_b=-\omega_m$ as shown in Ref.~\protect\cite{Barzanjeh}.} \label{fig1}
\end{figure}

In general the entanglement of the output modes is realized when the detunings, $\Delta_j$ with $j=a,b$, between the frequencies of the driving fields and that of the corresponding cavity modes have opposite sign, and coincide in modulus with the mechanical frequency $\omega_m$ \cite{Barzanjeh}.
In Fig.~\ref{fig1}, we assume this condition ($\Delta_a=-\Delta_b=\omega_m$), and we
study the logarithmic negativity between the two output modes  as a function of the central frequency $\Omega_a$ of the filter of the output mode A, when the frequency of the filter of the output mode B is fixed at $\Omega_b=\omega_m$, and for two different values of the
Q-factor of the mechanical resonator; Specifically the blue dotted curve corresponds to $Q_m=10^5$ and the red solid curve to $Q_m=10^7$.
These results demonstrate that the entanglement  between the two optical output modes is optimized when the filtered modes fulfill the resonance condition $\Omega_a=-\Omega_b=\omega_m$.
In general,  two-mode squeezing is observed for filtered modes for which $\Omega_a=-\Omega_b$~\cite{Zippilli14}, and the two-mode squeezing extends over a bandwidth of the order of the decay rates $\kappa_j$.
Hence, in order to resolve the corresponding entanglement properties, one has to select temporal filtered modes, whose duration, $\tau_j$, is much larger then $1/\kappa_j$.
In turn, the duration of the filtered modes determines the width of the curves in Fig.~\ref{fig1}, that is of the order of $1/\tau_j$.
Specifically Fig.~\ref{fig1} is obtained for $\tau_j\sim5\times10^{-6}\,$s, that corresponds to an optical length of the filtered mode of $\sim 1.5\,$Km.

Moreover, we observe that the logarithmic negativity is significantly sensitive to the mechanical quality factor, and it increases for increasing values of the $Q$.
We finally remark that the results of Fig.~\ref{fig1} demonstrate that a realistic optomechanical interaction is able to entangle efficiently fields at two significantly different frequencies, which could be exploited for establishing long distance quantum communication. In particular, Fig.~\ref{fig1} refers to the entanglement between one field at telecom wavelength (1550 nm) and another at a larger frequency, in the near-infrared (810 nm), falling within a transparency window for optical communications through the atmosphere. We shall see in the next Section that such ``hybrid'' entanglement could be exploited for circumventing the limitations due to photon losses in quantum communication along telecom fibers.

\section{Steady State Entanglement Swapping}

In general the distances over which it is possible to transmit entangled resources is limited by the unavoidable noise and losses affecting any transmission process. These limitations can be partially overcome, for example, by using multiple entanglement sources and swapping operations~\cite{zukowski,pan,xia,takei,lloyd,mehdi1,mehdi2,mehdi3}.
In this section we describe how two copies of the CV entangled light generated by two distant devices similar to the one described in the previous section can be used to extend the entanglement to larger distances. This can be realized by sending two outputs at the same frequency of the two devices to an intermediate location, where a Bell-measurement is performed. This approach, hence, realizes a trusted quantum relay~\cite{zoller,gisin}. The states of the two entangled sources are equal and they are characterized by covariance matrices of the form of Eq.~(\ref{cov}). We distinguish the two matrices using different indices, namely  $\bf V_{AC}$ and $\bf V_{BC}$ for the state generated form the first and second source
respectively. Two output fields, one from each source are sent to the distant locations at
Alice and Bob sites, respectively, denoted by the indices $A$ and $B$.
On the other hand, the remaining output fields are sent to Charlie, at an intermediate location, and denoted with the index $C$ in both matrices.
We now follow the standard procedure of entanglement swapping between continuous variables as described in~\cite{xia,takei,lloyd,mehdi1,mehdi2,mehdi3}
To be more specific, we introduce the following notation for the two covariance matrices
\begin{equation}
{\bf V_{AC}}=\left( \begin{array}{cc}
{\bf A}    &{\bf D}_1      \\
{\bf D}^T_1    &{\bf C_1}
\end{array}\right), \quad
{\bf V_{BC}}=\left( \begin{array}{cc}
{\bf B}    &{\bf D}_2      \\
{\bf D}^T_2    &{\bf C}_2
\end{array}\right),  \label{in1}
\end{equation}
however, keeping in mind that, in reality, ${\bf V_{AC}}$ and ${\bf V_{BC}}$ are assumed to be equal and the different symbols are used only to clarify the description of the swapping dynamics.

\begin{figure}
\includegraphics[scale=1]{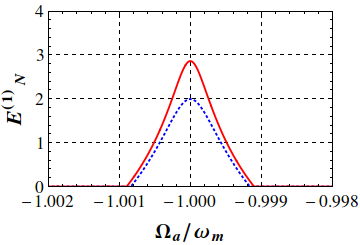}
\caption{(Color online) Entanglement between Alice and Bob
prepared by swapping the entanglement of two pairs of fields produced with equal optomechanical devices whose parameters are those of Fig.~\protect\ref{fig1}.} \label{fig5}
\end{figure}

In continuous variables, Bell measurements are realized with balanced homodyne detections. In the swapping protocol this is realized by Charlie on two modes at the same frequencies, whose quadrature operators are $\hat X_{out,b}^{(\ell)}$ and $\hat Y_{out,b}^{(\ell)}$, with $[\hat X_{out,b}^{(\ell)} , \hat Y_{out,b}^{(\ell')}]=i\,\delta_{\ell,\ell'}$, where the index $\ell=1,2$ distinguishes the fields resulting from two distant entangler devices.
In the present case we choose the two outputs at the near-infrared frequency so that at the end of the protocol the two modes at telecom frequency become entangled without ever having interacted.
The balanced homodyne detection performed by Charlie realizes the measurement of the commuting quadratures $(\hat X_{out,b}^{(1)}-\hat X_{out,b}^{(2)})$ and $(\hat Y_{out,b}^{(1)}+\hat Y_{out,b}^{(2)})$.
We denote the result of the corresponding measurements with the parameters $x_-$ and $y_+$ respectively.
After the Bell measurement, the two telecom fields are in a two-mode squeezed state with a non-zero mean field. The corresponding covariance matrix takes the form~\cite{lloyd,JMO,Pirandola&Mancini}
\begin{equation}
{\bf V_{AB}}=\left( \begin{array}{cc}
{\bf A}-{\bf D}_1{\bf Z}{\bf M^{-1}}{\bf Z}{\bf D}^T_1  &{\bf D}_1{\bf Z}{\bf M^{-1}}{\bf D}^T_2      \\
{\bf D}_2{\bf M^{-1}}{\bf Z}{\bf D}^T_1    &{\bf B} -  {\bf D}_2 {\bf M^{-1}} {\bf D}^T_2
\end{array}\right), \label{cov1}
\end{equation}
where ${\bf M}$ is the symmetric matrix ${\bf M}={\bf Z}{\bf C}_1{\bf Z}+{\bf C}_2$,  with ${\bf Z}=Diag(1,-1)$. Moreover the vector for the mean values of the field quadratures $\overrightarrow {\bf d}_{AB}=[\langle\hat X_{out,a}^{(1)}\rangle,\langle\hat Y_{out,a}^{(1)}\rangle,\langle\hat X_{out,a}^{(2)}\rangle,\langle\hat Y_{out,a}^{(2)}\rangle]^T$ (also called displacement or drift vector), is given by
\begin{equation}
\overrightarrow {\bf d}_{AB}=2
\left( \begin{array}{c}
-{\bf D}_1{\bf Z}{\bf M^{-1}}\ \overrightarrow {\bf k} \\
{\bf D}_2 {\bf M^{-1}}\  \overrightarrow {\bf k}
\end{array}\right),
\label{ch-dis}
\end{equation}
where $\overrightarrow {{\bf k}}=(x_-  , y_+)^T$ is the vector formed by the Bell measurement results. The entanglement of the state shared by Alice and Bob can be measured in terms of the logarithmic negativity, which is readily computed from the definition of Eq.~(\ref{EN}) applied to the covariance matrix $\bf V_{AB}$.
Specifically we can use Eq.~(\ref{cov1}) and the definition of $\bf M$ to express $\bf V_{AB}$ in terms of the elements of the matrix in Eq.~(\ref{cov}). The result is
\begin{equation}
{\bf V_{AB}}=\left( \begin{array}{cc}
{\bf V_{11}}    &{\bf V_{12}}      \\
{\bf V_{12}}     &{\bf V_{11}}   \label{first}
\end{array}\right)
\end{equation}
\noindent with
\begin{equation}
{\bf V_{11}}=\left( \begin{array}{cc}
({\bf V}_{out})_{11}-\frac{1}{2}[\frac{({\bf V}_{out})^2_{13}}{({\bf V}_{out})_{33}}+\frac{({\bf V}_{out})^2_{14}}{({\bf V}_{out})_{44}}] &({\bf V}_{out})_{12}-\frac{1}{2}[\frac{({\bf V}_{out})_{13}({\bf V}_{out})_{23}}{({\bf V}_{out})_{33}} +\frac{({\bf V}_{out})_{14}({\bf V}_{out})_{24}}{({\bf V}_{out})_{44}} ]   \\
({\bf V}_{out})_{12}-\frac{1}{2}[\frac{({\bf V}_{out})_{13}({\bf V}_{out})_{23}}{({\bf V}_{out})_{33}} +\frac{({\bf V}_{out})_{14}({\bf V}_{out})_{24}}{({\bf V}_{out})_{44}} ]    &({\bf V}_{out})_{22}-\frac{1}{2}[\frac{({\bf V}_{out})^2_{23}}{({\bf V}_{out})_{33}}+\frac{({\bf V}_{out})^2_{24}}{({\bf V}_{out})_{44}}]
\end{array}\right), \label{V11}
\end{equation}
\noindent and
\begin{equation}
{\bf V_{12}}=\left( \begin{array}{cc}
\frac{1}{2}[\frac{({\bf V}_{out})^2_{13}}{({\bf V}_{out})_{33}}-\frac{({\bf V}_{out})^2_{14}}{({\bf V}_{out})_{44}}]  &\frac{1}{2}[\frac{({\bf V}_{out})_{13}({\bf V}_{out})_{23}}{({\bf V}_{out})_{33}} -\frac{({\bf V}_{out})_{14}({\bf V}_{out})_{24}}{({\bf V}_{out})_{44}} ]   \\
\frac{1}{2}[\frac{({\bf V}_{out})_{13}({\bf V}_{out})_{23}}{({\bf V}_{out})_{33}} -\frac{({\bf V}_{out})_{14}({\bf V}_{out})_{24}}{({\bf V}_{out})_{44}} ]    &\frac{1}{2}[\frac{({\bf V}_{out})^2_{23}}{({\bf V}_{out})_{33}}-\frac{({\bf V}_{out})^2_{24}}{({\bf V}_{out})_{44}}]
\end{array}\right). \label{V12}
\end{equation}
These matrices can be then used to compute the symplectic eigenvalue as in Eq.~(\ref{eta}) and, hence, the logarithmic negativity. The corresponding results are shown in Fig.~\ref{fig5} where we observe that the entanglement is reduced with respect to that of the initial resources (see Fig.~\ref{fig1}) as expected for swapping operations realized with non-maximal entanglement. In particular, we note that, in the continuous variable case, the entanglement can never reach the maximum set by the infinitely squeezed state, since it corresponds to an unphysical situation with infinite energy~\cite{xia,takei,lloyd}.

\paragraph*{Teleportation Fidelity:}

The amount of swapped entanglement can also be characterized in terms of its effectiveness as a quantum channel usable for quantum teleportation. In particular, we can analyze the performance of the quantum channel realized by the swapping protocol in terms of the fidelity for the teleportation of a pure coherent state from Alice to Bob.
In CV teleportation protocols, Alice mixes on a beam splitter the state to be teleported with the part of the entangled state in her hand. Then she performs a Bell measurement on the fields, and communicates the corresponding results to Bob who has to perform a specific displacement operation, conditioned on the result of the measurement, in order to obtain the teleported state.
In practice, the state to be teleported is a signal of finite duration set by its coherence time $\tau_c$. Correspondingly all the detection processes are in reality done over correspondingly finite time intervals.
Therefore, in order for the stationary description of the system dynamics to be valid the coherence time of the signal have to be longer than the time needed by the entangler device to reach its steady state.

The corresponding teleportation fidelity is most easily described in terms of the Wigner's characteristic functions~\cite{JMO} of the channel $\Phi^{ch}(\mu_A,\mu_B)$ and of the input state to be teleported $\Phi^{in}(\mu)$.
To be specific
\begin{equation}
\Phi^{ch}(\mu_A,\mu_B)=e^{-\overrightarrow \mu_{AB}^T  {\bf V_{AB}}\overrightarrow \mu_{AB}+ i \overrightarrow {\bf d}^T_{AB} \overrightarrow \mu_AB}\ ,
\label{ch}
\end{equation}
where $\mu_A$ and $\mu_B$ are the complex phase-space variable for the two modes and $\overrightarrow \mu_{AB}$ is the vector of the corresponding real and imaginary parts defined as $\overrightarrow \mu_{AB}=({\rm Im}[\mu_A], -{\rm Re}[\mu_A] ,{\rm Im}[\mu_B], -{\rm Re}[\mu_B] )^T$;
furthermore the characteristic function for the input state, with covariance matrix ${\bf V_{in}}$ and drift vector $\overrightarrow d_{in}$, is
\begin{equation}
\Phi^{in}(\mu)=e^{-\overrightarrow \mu^T {\bf V_{in}} \overrightarrow \mu + i \overrightarrow d^T_{in} \overrightarrow \mu}, \label{in}
\end{equation}
where now the phase space variable is $\mu$ and $\overrightarrow \mu=({\rm Im}[\mu], -{\rm Re}[\mu] )^T$.
Specifically, here we consider the teleportation of a coherent state for which ${\bf V_{in}}=Diag(1/2,1/2)$.
The corresponding teleportation fidelity averaged over all the possible measurement results obtained by Alice, is then given by
 \cite{JMO}
\begin{equation}
F(\delta)=\frac{1}{\pi}\int d^2\mu |\Phi^{in}(\mu)|^2 [\Phi^{ch}(-\mu^*,\mu)]^* e^{\delta \mu^*-\delta^* \mu}\ .\label{fid}
\end{equation}
Here the quantity $\delta$ accounts for the displacement operation that Bob has to perform on his side to balance the channel displacement~\cite{JMO}. In our case this is due to the drift induced by the Bell measurements of the swapping protocol, that is defined in Eq.~(\ref{ch-dis}), and which depend upon the values $x_-$ and $y_+$ of the homodyne measurements performed by Charlie.
In order for Bob to correctly balance this displacement he should have knowledge of $x_-$ and $y_+$.
These values are therefore communicated to Bob by Charlie over a classical channel.
Once the correct displacement is performed by Bob, the resulting teleportation fidelity becomes~\cite{lloyd,JMO,fiurasek}
\begin{equation}
F=	\frac{1}{\sqrt{Det {\bf\Gamma}}} \label{fin}
\end{equation}
\noindent with
\begin{equation}
{\bf\Gamma} = 2{\bf V_{in}} +{\bf A}+{\bf Z}{\bf A}{\bf Z}-({\bf C}{\bf Z}+{\bf Z}{\bf C}), \label{tot}
\end{equation}
where we have used the fact that the two entangler devices are equal and, therefore, ${\bf A}={\bf B}$ and ${\bf C_1}={\bf C_2}={\bf C}$.
\begin{figure}
\includegraphics[scale=1]{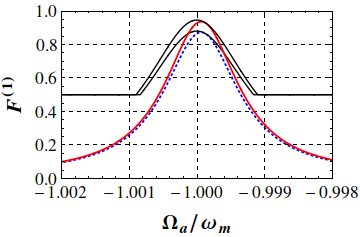}
\caption{(Color online)
Teleportation fidelity of a coherent state in the case when Alice and Bob share an entangled quantum channel prepared via entanglement swapping of two pairs of fields produced with equal optomechanical devices with the parameters given in Fig.~\protect\ref{fig1}.
The black curves represent the value of the bound defined in Eq.(\protect\ref{bound}).
} \label{fig2}
\end{figure}

\begin{figure}
\includegraphics[scale=1]{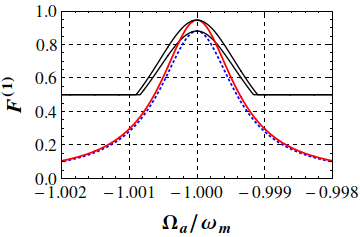}
\caption{(Color online) Teleportation fidelity as in Fig.~\protect\ref{fig2}, but after optimization over local field rotations.
The values of the parameters are the same as in Fig.~\protect\ref{fig1}. } \label{fig4}
\end{figure}

In Fig.~\ref{fig2} we represent the fidelity of the teleported initial coherent state as a function of the frequency of the fields filtered from the near-infrared output modes,
and at fixed values, $\Omega_b=\omega_m$, of the frequency of the fields filtered from the modes at telecom wavelength. We observe that the maximum of the fidelity is slightly shifted with respect to the position of the maximum of the entanglement, which is found at $\Omega_a=-\omega_m$. This behaviour is due to the non perfect choice of the combination of the modes measured by Alice in the teleportation protocol. In any case, the fidelity respect the bound defined in Ref.~\cite{Mari&David,Adesso2005}, that is given by
\begin{equation}
F = \frac{1}{1+e^{-E_N}},
\label{bound}
\end{equation}
where $E_N$ is the logarithmic negativity of the quantum channel.
We finally remark that the coincidence between maximum entanglement and maximum fidelity can be attained by
performing appropriate local rotations of the field quadratures at Alice's and Charlie's sites.
We note that being local these rotations do not affect the entanglement, but only the fidelity.
The maximum at $\Omega_a=-\omega_m$ is recovered by maximizing the fidelity over local rotations, as it is shown in Fig. \ref{fig4}. Together with the coincidence of the maxima, we observe that the fidelity reaches the maximum value set by the bound in Eq.~(\ref{bound}).

\section{Continuous Variable Concatenated Swaps}

Quantum communications over longer distances can be realized with a series of concatenated swaps over many entangled pairs~\cite{sanders}.
However, as we have already seen, every swap operation reduces the amount of entanglement. This imposes a trade-off between maximum swapped entanglement and maximum attainable distance with this kind of system.
Furthermore, in general, the amount of resources needed for relay-based large-distance quantum communications scales exponentially with the distance \cite{GisinRMP}.
Notwithstanding, in the lack of CV quantum repeaters this approach is surely better than the short distance bound imposed by CV quantum protocols~\cite{Grangier&Diamanti}.

Here we consider a series of equal optomechanical devices able to produce several entangled field pairs usable for concatenated swapping operations and we analyze the performance in generating long-distance entanglement.
These devices have to be arranged as in Fig.~\ref{fig5}, where now the characters are Alice, Bob and three Charlies. In particular, each pair of modes used to perform the intermediate Bell measurements needs to have the same frequency and, correspondingly, the end fields at Alice and Bob sites are two fields at telecom frequency.
We consider a specific example consisting of three swaps of the kind discussed in the previous Section, realized by the three Charlies halfway between two successive devices.
The state of each pair of filtered fields generated by the four devices is described by a covariance matrix equal to Eq.~(\ref{cov}). Here we include an index $\ell\in\{1,2,3,4\}$ to distinguish the four entangled pairs so that the initial covariance matrices can be expressed as
\begin{equation}
{\bf V_{\ell}}=\left( \begin{array}{cc}
{\bf A_\ell}    &{\bf D}_\ell      \\
{\bf D}^T_\ell   &{\bf C_\ell}
\end{array}\right),
\label{in4}
\end{equation}
keeping in mind, however, that we assume that they are all equal.
\begin{figure}
\includegraphics[scale=.4]{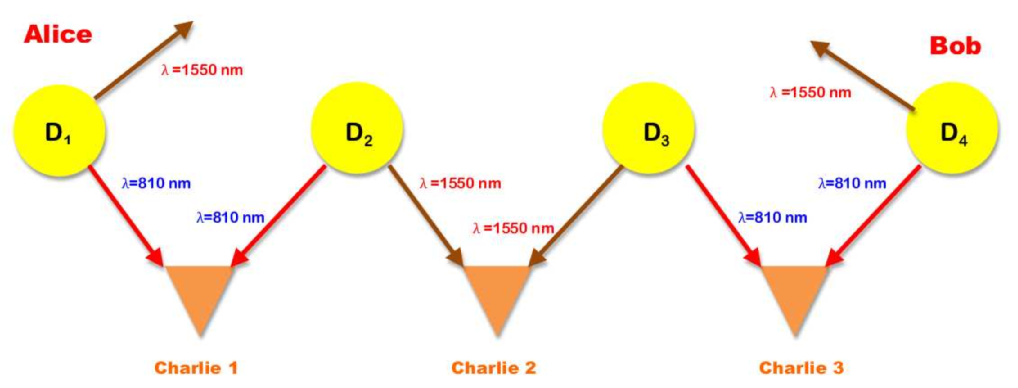}
\caption{(Color online) Chain of concatenated entangler devices, $D_j$ for $j=1,2,3,4$, each of which is equal to that discussed in Sec.~\ref{model}. Each Charlie performs swap operations over two fields at the same frequency. At the end of the concatenated swap process, Alice and Bob share two entangled fields at telecom frequencies.
 \label{figscheme}}
\end{figure}
After the first swap performed by Charlie 1 (see Fig.~\ref{figscheme})
the field at Alice's site gets correlated with one of the fields of Charlie 2, and they are described by the covariance matrix
\begin{equation}
{\bf V_{1,2}}=\left( \begin{array}{cc}
{\bf A}_1-{\bf D}_1{\bf Z}{\bf M_1^{-1}}{\bf Z}{\bf D}^T_1  &{\bf D}_1{\bf Z}{\bf M_1^{-1}}{\bf D}^T_2      \\
{\bf D}_2{\bf M_1^{-1}}{\bf Z}{\bf D}^T_1    &{\bf A}_2 -  {\bf D}_2 {\bf M_1^{-1}} {\bf D}^T_2
\end{array}\right) =\left( \begin{array}{cc}
{\bf A}_{sw}^{(1)}    &{\bf D}_{sw}^{(1)}      \\
{\bf D}_{sw}^{(1)T}    &{\bf C}_{sw}^{(1)}
\end{array}\right),
 \label{1sw}
\end{equation}
with ${\bf M_1}={\bf Z}{\bf C}_1{\bf Z}+{\bf C}_2={\bf M_1}^T$ and
which, of course, coincide with Eq.~(\ref{cov1}). After the second swap we get similarly the covariance matrix for the correlated fields shared by Alice and Charlie 3
\begin{equation}
{\bf V_{1,3}}=\left( \begin{array}{cc}
{\bf A}_{sw}^{(1)}-{\bf D}_{sw}^{(1)}{\bf Z}{\bf M_2^{-1}}{\bf Z}{\bf D}_{sw}^{(1)T}  &{\bf D}_{sw}^{(1)}{\bf Z}{\bf M_2^{-1}}{\bf D}^T_3      \\
{\bf D}_3{\bf M_2^{-1}}{\bf Z}{\bf D}_{sw}^{(1)T}    &{\bf A}_3 -  {\bf D}_3 {\bf M_2^{-1}} {\bf D}^T_3
\end{array}\right) =\left( \begin{array}{cc}
{\bf A}_{sw}^{(2)}    &{\bf D}_{sw}^{(2)}      \\
{\bf D}_{sw}^{(2)T}    &{\bf C}_{sw}^{(2)}
\end{array}\right),
 \label{2sw}
\end{equation}
with ${\bf M_2}={\bf Z}{\bf C}_{sw}^{(1)}{\bf Z}+{\bf C}_3={\bf M_2}^T$, and
finally, after the third swap, the remaining fields at Bob's and Alice's sites are described by the matrix
\begin{equation}
{\bf V_{1,4}}=\left( \begin{array}{cc}
{\bf A}_{sw}^{(2)}-{\bf D}_{sw}^{(2)}{\bf Z}{\bf M_3^{-1}}{\bf Z}{\bf D}_{sw}^{(2)T}  &{\bf D}_{sw}^{(2)}{\bf Z}{\bf M_3^{-1}}{\bf D}^T_4      \\
{\bf D}_4{\bf M_3^{-1}}{\bf Z}{\bf D}_{sw}^{(2)T}    &{\bf A}_4 -  {\bf D}_4 {\bf M_3^{-1}} {\bf D}^T_4
\end{array}\right).
 \label{3sw}
\end{equation}
with ${\bf M_3}={\bf Z}{\bf C}_{sw}^{(2)}{\bf Z}+{\bf C}_4={\bf M_3}^T.$
The corresponding logarithmic negativity is shown in Fig. \ref{fig6}. It demonstrates a significant amount of entanglement shared by Alice and Bob, and correspondingly a potentially usable quantum Gaussian channel.
\begin{figure}
\includegraphics[scale=1]{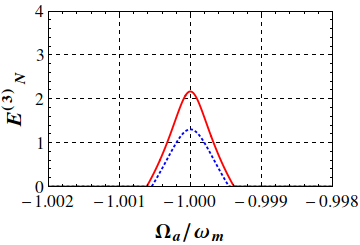}
\caption{(Color online) Logarithmic negativity $E^{(3)}_{N}$ between Alice and Bob, after three intermediate swaps, with $Q_m=10^7$ in all devices (red curve), and $Q_m=10^5$(blue dotted curve). The values of the other parameters are the same as in Fig. \ref{fig1}. } \label{fig6}
\end{figure}
As in the previous section, its efficiency can be characterized in terms of the teleportation fidelity.
We remark that in this case, in order to balance the channel drift induced by the swap operations, Bob has to take into account the results of all the intermediate measurements.
The corresponding fidelity for the teleportation of a coherent state optimized over the local rotations, as discussed in the previous section, is reported in Fig. \ref{fig7}.
\begin{figure}
\includegraphics[scale=1]{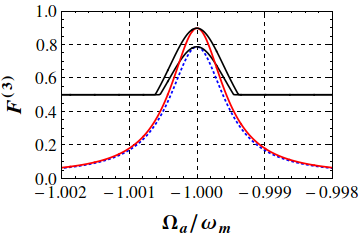}
\caption{(Color online) Teleportation fidelity of an unknown coherent state, optimized over local rotations,
with $Q_m=10^7$ (red curve), and $Q_m=10^5$ (blue dotted curve) in all devices. The black solid curves represent the fidelity bounds defined in Eq.~(\ref{bound}). The values of the other parameters are the same as in Fig.~\ref{fig1}. } \label{fig7}
\end{figure}

\section{Effect of losses in the CV Concatenated Swaps}

So far we have considered no losses during transmission between the various nodes involved in the concatenated swap protocol, and unit detection efficiencies. In reality, the detection is never perfect and any transmission process is affected by losses of various physical nature, which result in the attenuation of the transmitted signal.

Here we take into account detection inefficiencies and losses of the transmission channels, of length $\it l$ between the entangler devices and the detection sites, in terms of the phenomenological quantum efficiency $\eta=\eta_0 e^{-\alpha \it{l}}$, where $\eta_0$ is the homodyne detection efficiency of all the measurement devices used by the three Charlies, and $\alpha$ the transmission loss in ${\rm dB/km}$~\cite{sanders}.
We note also that the effect of the losses can be modeled as an additional beam splitter on the transmission channel with vacuum noise input in the second port.

\begin{figure} \includegraphics[scale=1]{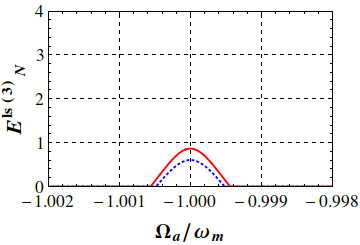}
\caption{(Color online) Logarithmic negativity $E_N$  after three swaps, in presence of loss with $\alpha=0.005 {\rm dB/km }$ and homodyne efficiency $\eta_0=.95$ at each Charlie's measurement, with $Q_m=10^7$ (red curve) and $Q_m=10^5$ (blue dotted curve). The values of the other parameters are the same as in Fig.~\ref{fig1}.} \label{fig8}
\end{figure}

\begin{figure}
\includegraphics[scale=1]{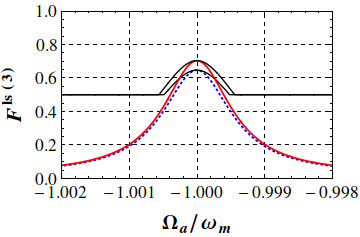}
\caption{(Color online) Teleportation fidelity of an unknown coherent state after three swaps, for $\alpha=0.005 {\rm dB/km} $ and $\eta_0=0.95$ at each Charlie's measurement, with $Q_m=10^7$ (red curve) and $Q_m=10^5$ (blue dotted curve). The values of the other parameters are the same as in Fig.~\ref{fig1}.}  \label{fig9}
\end{figure}

\begin{figure}
\includegraphics[scale=1]{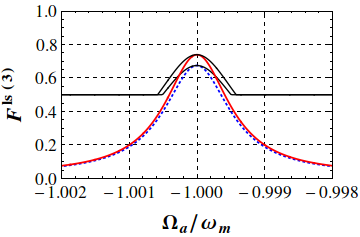}
\caption{(Color online) Teleportation fidelity of an unknown coherent state after three swaps, for $\alpha=0.005 {\rm dB/km} $ and $\eta_0=0.99$ at each Charlie's measurement, with $Q_m=10^7$ (red curve) and $Q_m=10^5$ (blue dotted curve). The values of the other parameters are the same as in Fig.~\ref{fig1}.}  \label{fig10}
\end{figure}

Assuming equal losses for all the fields, their effect on the covariance matrices is simply expressed by the formula
\begin{eqnarray}\label{Vellls}
{\bf V_\ell^{ls}}= \eta\ {\bf V_\ell}\ +\mathbb{1}
\frac{1-\eta}{2} 
\end{eqnarray}
where $\bf V_\ell$ is defined in Eq.~(\ref{in4}) and $\mathbb{1}$ is the identity matrix. The concatenated swapping protocol remains, therefore, equal to that described in the preceding section, but now it is applied to the matrices in Eq.~(\ref{Vellls}).
In this case the corresponding logarithmic negativity of the generated end-to-end entanglement and the teleported fidelity depend upon
the detection efficiency $\eta_0$ and the loss parameter $\alpha$.
According to Ref.~\cite{bams}, in the case of extremely clear day with visibility between 50-150 km, the loss in free space varies between 0.04 and 0.005 ${\rm dB/km}$ for a laser with 850 nm or 1550 nm central frequency. Assuming the best value and considering the free space between the devices to be 40 km and a homodyne quantum efficiency of $\eta_0=0.95$, we find the curves reported in Fig.~\ref{fig8} for the logarithmic negativity, and in Fig.~\ref{fig9} for the optimized teleportation fidelity of a coherent state.
In this case we see the dramatic effect of $\eta$ especially on the fidelity which, in the case of low Q-factor, goes below the limit for the teleportation of nonclassical states, which is set at $2/3$~\cite{Ban}. For a fixed distance $l$, the only way to increase the teleportation fidelity is to enhance the homodyne detection efficiency; in particular if we set the detection efficiency
at $\eta_0=0.99$, as shown in Fig.~\ref{fig10}, the transmission fidelity is significantly increased and exceeds the $2/3$ limit.

 \section{Concluding Remarks}

We have shown that the entanglement generated with optomechanical systems could be efficiently exploited for quantum communications between distant sites. Specifically we have seen that consecutive swap operations performed over a series of entangled optical fields pairs, generated with optomechanical systems, may serve to distribute entanglement over large distances. Such generated entanglement can be used as a channel for efficient quantum communications. We have characterized its efficiency in terms of the corresponding fidelity for the teleportation of coherent states, showing that using mechanical resonators with very high Q-factor it is possible to extend the quantum communication of telecom stationary signals up to hundreds of kilometers.
Such distances are mainly limited by the homodyne detection efficiency, by the photon loss along the transmission channel, and by the quality factor of mechanical device which tend to degrade the initially available entanglement provided by the cavity optomechanical source, which sets an upper bound to the fidelity of a teleported state~\cite{Mari&David,Adesso2005}.
Therefore, in order to increase the fidelity, and possibly the distance of the communication, one should improve the entanglement by some means.
In principle this can be achieved by employing more elaborated strategies that include distillation stages~\cite{dist,eisert2} at each swap.

A possible way to implement the proposed concatenated swap scheme is to put Charlie performing the Bell measurement of the swapping protocol on a low-orbit satellite, and the entangler devices on distant high mountain telescopes. In this condition, the entangled components at the near-infrared wavelength within the atmosphere transparency window can be efficiently transmitted from the Earth to the satellite through the atmosphere, whose effective length can be estimated to be around 20 Km.
The other components of the entangled fields are instead sent over telecom fibers to the distant Alice and Bob, that perform the teleportation protocol.
In this scenario the technological difficulties would be time synchronization, satellite tracking and a sufficiently large detection area on the satellite. However these issues do not seem more problematic than those usually encountered in classical Earth-satellite communications.
Similar considerations have been addressed in Ref.~\cite{Weinfurter}, where quantum key distribution has been realized between an airplane and a ground station using advanced pointing and tracking systems.


\begin{thebibliography}{99}


\bibitem{Eckert} A. K. Ekert Phys. Rev. Lett. {\bf 67}, 661 (1991).

\bibitem{Yamamoto} E. Waks, A. Zeevi, and Y. Yamamoto Phys. Rev. A, {\bf 65}, 52310 (2002).



\bibitem{Bennett}  C. H. Bennett {\it et al.} Phys. Rev. Lett. {\bf 70}, 1895 (1993).

\bibitem{Braunstein&Kimble}S. L. Braunstein and H. J. Kimble, Phys. Rev. Lett.{\bf 80}, 869 (1998).

\bibitem{Bennett2}C. H. Bennett and S. J. Wiesner, Phys. Rev. Lett. {\bf 69},  2881 (1992).

\bibitem{Braunstein&Kimble2}S. L. Braunstein and H. J. Kimble, Phys. Rev. A {\bf 61}, 042302 (2000).

\bibitem{kimble} H. J. Kimble, Nature {\bf 453}, 1023 (2008).

\bibitem{Aspel} M. Aspelmeyer, {\it et al.} Science {\bf 301}, 621 (2003).

\bibitem{zeilinger} Xiao-song Ma,  {\it et al.} Nature {\bf 489}, 269 (2012).

\bibitem{cinesi} Juan Yin, {\it et al.}  Nature {\bf 488}, 185 (2012).

\bibitem{space} R. Ursin {\it et al.; IAC Microgravity Sciences and Processes Symposium} Glasgow, Scotland UK  2008.

\bibitem{zoller} H. J. Briegel, {\it et al.}  Phys. Rev. Lett. {\bf 81}, 5932 (1998); L. M. Duan,  {\it et al.}  Nature {\bf 414}, 413 (2001).

\bibitem{gisin} N. Gisin, {\it et al.} , Rev. Mod. Phys. {\bf 74}, 145 (2002).

\bibitem{Eisert2003}
J. Eisert and M. B. Plenio,
Int. J. Quantum Inform. {\bf 01}, 479 (2003).

\bibitem{Braunstein2005}
S. L. Braunstein and P. van Loock,
Rev. Mod. Phys. {\bf 77}, 513 (2005).

\bibitem{Paris2005}
A. Ferraro, S. Olivares and M. G. A. Paris, {\it Gaussian States in Quantum Information} (Bibliopolis, Napoli, 2005).

\bibitem{Adesso2007}
G. Adesso and F. Illuminati,
J. Phys. A: Math. Theor. {\bf 40}, 7821 (2007).

\bibitem{Weedbrook20012}
C. Weedbrook, S. Pirandola, R. Garc\'ia-Patr\'on, N. J. Cerf, T. C. Ralph, J. H. Shapiro, and S. Lloyd,
Rev. Mod. Phys. {\bf 84}, 621 (2012).

\bibitem{AMO} C. Genes, A. Mari, D. Vitali, and P. Tombesi, {\it Advances in Atomic, Molecular, and Optical Physics} {\bf 57}, 33 (2009)

\bibitem{rmp} M. Aspelmeyer, T. J. Kippenberg, and F. Marquardt, Rev. Mod. Phys. \textbf{86}, 1391 (2014).

\bibitem{oconnell}A. D. O'Connell \textit{et al.}, Nature (London)\textbf{464}, 697 (2011).

\bibitem{teufel}J. D. Teufel \textit{et al.}, Nature (London)\textbf{475}, 359 (2011).

\bibitem{chan}J. Chan \textit{et al.}, Nature (London)\textbf{478}, 89 (2011).

\bibitem{brooks}D. W. C. Brooks, T. Botter, S. Schreppler, T. P. Purdy, N. Brahms, and D. M. Stamper-Kurn, Nature (London), \textbf{448}, 476 (2012).
\bibitem{painter} A. H. Safavi-Naeini, S. Gr\"oblacher, J. T. Hill, J. Chan, M. Aspelmeyer, O, Painter, Nature \textbf{500}, 185 (2013).
\bibitem{regal} T. P. Purdy, P-L. Yu, R. W. Peterson, N. S. Kampel, and C. A. Regal, Phys. Rev. X \textbf{3}, 031012 (2013).

\bibitem{verhagen}E. Verhagen,	S. Del\`eglise,	S. Weis, A. Schliesser, and T. J. Kippenberg, Nature \textbf{482}, 63 (2012).
\bibitem{palomaki}T. A. Palomaki, J. W. Harlow, J. D. Teufel, R. W. Simmonds and K. W. Lehnert, Nature \textbf{495}, 210 (2013).
\bibitem{Lenhert} T. A. Palomaki {\it et al.}, Science, \textbf{342}, 710 (2013).
\bibitem{giovannetti} V. Giovannetti, S. Mancini, and P. Tombesi, Europhys. Lett. \textbf{54}, 559 (2001).

\bibitem{claudiu&mari} C. Genes, A. Mari, P. Tombesi, and  D. Vitali Phys. Rev. A \textbf{78}, 032316 (2008).

\bibitem {shabir1}Sh. Barzanjeh, D. Vitali, P. Tombesi, and G. J. Milburn, Phy. Rev. A \textbf{84}, 042342 (2011).

\bibitem{lin} L. Tian, Phys. Rev. Lett. {\bf 108}, 153604 (2012).

\bibitem{clerk} Y-D. Wang and A. A. Clerk, Phys. Rev. Lett. {\bf 108}, 153603 (2012).

\bibitem{Barzanjeh} Sh. Barzanjeh, M. Abdi, G. J. Milburn, P. Tombesi, and  D. Vitali Phys. Rev. Lett. {\bf 109}, 130503 (2012).

\bibitem{hill}J. T. Hill, A. H. Safavi-Naeini, J. Chan, and O. Painter, Nature Commun. \textbf{3}, 1196 (2012).

\bibitem{bochmann}J. Bochmann \textit{et al}., Nature Physics. \textbf{9}, 712 (2013).

\bibitem{bagci}T. Bagci \textit{et al}., Nature (London) \textbf{507}, 81 (2014).

\bibitem{lehnert14} R. W. Andrews, R.W. Peterson, T.P. Purdy, K. Cicak, R.W. Simmonds, C.A. Regal, and K.W. Lehnert, Nature Physics \textbf{10}, 321 (2014).

\bibitem{Grosshans&Grangier} F. Grosshans and Ph. Grangier, Phys. Rev. Lett. \textbf{88}, 057902 (2002).

\bibitem{zukowski} M. Zukowski, A. Zeilinger, M. A. Horne, and A. K. Ekert Phys. Rev. Lett. {\bf 71}, 4287 (1993).

\bibitem{pan}J.-W. Pan, D. Bouwmeester, H. Weinfurter, and A. Zeilinger, Phys. Rev. Lett. \textbf{80}, 3891 (1998).
\bibitem{xia} X. Jia, X. Su, Q. Pan, J. Gao, C. Xie, and K. Peng, Phys. Rev. Lett. \textbf{93}, 250503 (2004).
\bibitem{takei} N. Takei, H. Yonezawa, T. Aoki, and A. Furusawa, Phys. Rev. Lett. \textbf{94}, 220502 (2005).

\bibitem{lloyd} S. Pirandola, D. Vitali, P. Tombesi, and S. Lloyd, Phys. Rev. Lett. \textbf{97}, 150403 (2006).
\bibitem{mehdi1} M. Abdi, S. Pirandola, P. Tombesi, and D. Vitali, Phys. Rev. Lett. \textbf{109}, 143601 (2012).
\bibitem{mehdi2} M. Abdi, S. Pirandola, P. Tombesi, and D. Vitali, Phys. Rev. A \textbf{89}, 022331 (2012).
\bibitem{mehdi3} M. Abdi, P. Tombesi, and D. Vitali, Ann. Phys. (Berlin) \textbf{527}, 139 (2015).

\bibitem{Collett&Gardiner} M. J. Collett and C. W. Gardiner Phys. Rev. A {\bf 31 }, 3761 (1985).

\bibitem{Gardiner&Zoller} C. W. Gardiner and P. Zoller, {\it Quantum Noise} (Springer, Berlin, 2000).

\bibitem{Zippilli14} S. Zippilli, G. Di Giuseppe, and D. Vitali, accepted for publication in New J. Phys. (arXiv:1411.5609).

\bibitem{routh} M. Gopal {\it Control Systems: Principles and Design}, (Tata McGraw-Hill Education 2002).

\bibitem{Vidal&Werner} G. Vidal and R. F. Werner, Phys. Rev. A {\bf 65}, 032314 (2002).

\bibitem{eisert} J. Eisert, Ph.D. thesis, University of Potsdam, (2001).

\bibitem{plenio}M.B. Plenio, Phys. Rev. Lett. \textbf{95}, 090503 (2005).

\bibitem{illuminati} G. Adesso, A. Serafini, and F. Illuminati, Phys. Rev. A {\bf 70}, 022318 (2004).

\bibitem{JMO} S. Pirandola, S. Mancini, D. Vitali, and P. Tombesi, J. Mod. Opt. {\bf 51}, 901 (2004).

\bibitem{Pirandola&Mancini} S. Pirandola, and S. Mancini, Laser Physics {\bf 16}, 1418 (2006).

\bibitem{fiurasek} J. Fiurasek, Phys. Rev. A {\bf 66}, 012304 (2002).

\bibitem{Mari&David} A. Mari, and  D. Vitali Phys. Rev. A {\bf 78}, 062340 (2008).

\bibitem{Adesso2005}
G. Adesso and F. Illuminati,
Phys. Rev. Lett. {\bf 95}, 150503 (2005).

\bibitem{sanders} A. Khalique, W. Tittel, and B. C. Sanders, Phys. Rev. A {\bf 88}, 022336 (2013).

\bibitem{GisinRMP} N. Gisin, G. Ribordy, W. Tittel, and H. Zbinden, Rev. Mod. Phys. {\bf 74}, 145 (2002).

\bibitem{Grangier&Diamanti} P. Jouguet, S. Kunz-Jacques, A. Leverrier, Philippe Grangier, and E. Diamanti, Nature Phot. {\bf 7}, 378 (2013).

\bibitem{bams}
K. W. Fischer, M. R. Witiw, J. A. Baars, and T. R. Oke, 
Bull. Amer. Meteor. Soc. {\bf 85}, 725 (2004).


\bibitem{Ban}F. Grosshans and P. Grangier, Phys. Rev. A \textbf{64}, 010301(R) (2001); M. Ban, Phys. Rev. A \textbf{69}, 054304 (2004).

\bibitem{dist} C. H. Bennett {\it et al.} Phys. Rev. Lett. {\bf 76}, 722 (1996).

\bibitem{eisert2}J. Eisert,D. E. Browne, S. Scheel, and M. B. Plenio, Ann. Phys. \textbf{311} 431 (2004).

\bibitem{Weinfurter}S. Nauerth, F. Moll, M. Rau, Ch. Fuchs, J. Horwath, S. Frick, and H. Weinfurter, Nature Phot. {\bf 7}, 382 (2013)



\end{thebibliography}
\end{document}